\documentclass[a4paper,fleqn,usenatbib]{mnras}

\usepackage{newtxtext,newtxmath}

\usepackage[T1]{fontenc}
\usepackage{ae,aecompl}

\usepackage{graphicx}	
\usepackage{amsmath}	
\usepackage{amssymb}	

\usepackage{bm}
\usepackage[shortlabels]{enumitem}
\usepackage{caption}
\usepackage{subcaption}
\usepackage{soul}

\title[MHD instabilities of oscillating flows]{On Kelvin--Helmholtz and parametric instabilities driven by coronal waves}

\author[A. Hillier et. al.]{
Andrew Hillier$^{1}$\thanks{E-mail: a.s.hillier@exeter.ac.uk (AH)},
Adrian Barker$^{2}$,
I\~nigo Arregui$^{3,4}$,
Henrik Latter$^5$ 
\\
$^{1}$Department of Mathematics, CEMPS, University of Exeter, Exeter, EX4 4QF U.K. \\
$^{2}$Department of Applied Mathematics, School of Mathematics, University of Leeds, Leeds, LS2 9JT, U.K. \\
$^{3}$Instituto de Astrof\'{i}sica de Canarias, V\'{i}a L\'{a}ctea s/n, E-38205 La Laguna, Tenerife, Spain\\
$^{4}$Departamento de Astrof\'{i}sica, Universidad de La Laguna, E-38206 La Laguna, Tenerife, Spain\\
$^{5}$Department of Applied Mathematics and Theoretical Physics, University of Cambridge, Wiberforce Road, CB3 0WA U.K.\\
}

\date{Accepted XXX. Received YYY; in original form ZZZ}

\pubyear{2018}

\begin{document}
\label{firstpage}
\pagerange{\pageref{firstpage}--\pageref{lastpage}}
\maketitle

\begin{abstract}
The Kelvin--Helmholtz instability has been proposed as a mechanism to extract energy from magnetohydrodynamic (MHD) kink waves in flux tubes, and to drive dissipation of this wave energy through turbulence. It is therefore a potentially important process in heating the solar corona. However, it is unclear how the instability is influenced by the oscillatory shear flow associated with an MHD wave. We investigate the linear stability of a discontinuous oscillatory shear flow in the presence of a horizontal magnetic field within a Cartesian framework that captures the essential features of MHD oscillations in flux tubes. We derive a Mathieu equation for the Lagrangian displacement of the interface and analyse its properties, identifying two different instabilities: a Kelvin--Helmholtz instability and a parametric instability involving resonance between the oscillatory shear flow and two surface Alfv\'{e}n waves. The latter occurs when the system is Kelvin--Helmholtz stable, thus favouring modes that vary along the flux tube, and as a consequence provides an important and additional mechanism to extract energy. When applied to flows with the characteristic properties of kink waves in the solar corona, both instabilities can grow, with the parametric instability capable of generating smaller scale disturbances along the magnetic field than possible via the Kelvin--Helmholtz instability.
The characteristic time-scale for these instabilities is $\sim 100$\,s, for wavelengths of $200$\,km. The parametric instability is more likely to occur for smaller density contrasts and larger velocity shears, making its development more likely on coronal loops than on prominence threads.
\end{abstract}


\section{Introduction}\label{INTRO}

Recent observations of oscillating prominence threads by the Interface Region Imaging Spectrometer \citep[IRIS;][]{DEPON2014} and the Hinode Solar Optical Telescope \citep{TSU2008} show that they can `fade out' in cool lines (\ion{Ca}{II} and \ion{Mg}{II}) whilst simultaneously emerging in hotter lines (\ion{Si}{IV}) \citep{OKA2015}.
This could be a signature of heating in these structures.
\citet{OKA2015} compared their oscillations in the plane-of-sky motion (measured from \ion{Ca}{II} broadband images) and in the line-of-sight velocity field (from Dopplershifts in the \ion{Mg}{II} K line using spectra from IRIS); they found relative phase shifts of these oscillations between $90\deg$ and $180\deg$.
Forward modelling of simulated data showed that these observed relative phase shifts are consistent with the resonant absorption of a magnetohydrodynamic (MHD) {kink wave} \citep{ANTOLIN2015}.

{A key component of the model of \citet{ANTOLIN2015} is that the surface of an oscillating flux-tube can become unstable to the Kelvin--Helmholtz (KH) instability \citep[first seen associated with transverse kink waves in the simulations of][]{TERR2008}, a shear-flow instability common in astrophysical systems.}
Instances of the occurrence of this instability include: the interaction of the solar wind with the flanks of the magnetosphere \citep[e.g.][]{HASE2004}, erupting regions \citep{OFMAN2011}, the flanks of coronal mass ejections \citep{FOU2011, MOSTL2013}, and where emerging magnetic flux interacts with prominences \citep[e.g.][]{BERG2010, RYU2010, BERG2017}.
The KH instability can drive turbulence and is a potential way to dissipate the energy of surface Alfv\'{e}n waves in magnetic flux tubes \citep{HOLL1988, OFMAN1994, TERR2008, ANTOLIN2015}. {So the key question is: How do these shear flows develop at the surface of an oscillating flux tube?}

{The existence of an unstable shear flow at the surface of a flux tube undergoing a transverse kink wave can be understood from the eigenfunction of the linear wave \citep[e.g.][]{SAKU1991, GOOSSENS1992, GOOS2009}.
In the case where the density is discontinuous at the surface of the tube, a discontinuous shear flow exists there, but this becomes smooth for a continuous density profile \citep[e.g.][]{GOOS2009}.
However, for the smooth profile this shear flow can be enhanced by a process known as resonant absorption.
First proposed by \citet{ION1978}, resonant absorption occurs because of a resonance between a kink wave travelling along a flux tube and an Alfv\'{e}n wave, which leads to a velocity singularity at the tube surface in ideal MHD \citep{SAKU1991, GOOSSENS1992}, though non-ideal effects make this shear-flow velocity finite but large \citep{GOOSSENS1995}.
These small-scale flows enhance the dissipative processes \citep{HOLL1978, WEN1974, WEN1978, WEN1979}.
Recent theoretical and numerical work has been devoted to transverse kink waves \citep{GOOS2009}, 
and has shown that their non-linear dynamics develop Kelvin--Helmholtz (KH) unstable flows \citep{TERR2008, ANT2014, ANTOLIN2015, ANT2016, ANT2017, TERR2018}. 
The instability acts to extract energy from the large-scale mode and to distribute it to smaller scales where dissipation can act effectively. 
The cause and nature of this eventual heating, however, is still under investigation \citep{MAGYAR2016, HOW2017,KARA2017, ANT2018}.
For a recent review of wave-based heating in the solar atmosphere see \citet{ARREGUI2015}.}

The general mechanism of the KH instability is that it breaks up the shear layer at the boundary between two flows by creating vortices \citep{CHAN1961}. This may lead to turbulence via secondary 3D instabilities. For magnetohydrodynamic flows, magnetic tension works to suppress the KH instability and favours unstable modes that do not vary along the field.
To understand the growth of the magnetic KH instability in coronal loops, \citet{SOLER2010} investigated how it develops on the surface of a rotating flux tube. They found that the physics of the linear instability are not greatly altered by the change in geometry.
However, the influence of oscillations in the shear (as occurs in an MHD wave) on the growth of the instability is yet to be understood.

Oscillatory shear flows have been well studied in hydrodynamics.
\citet{KELLEY1965} investigated the instability of an oscillating shear flow including gravity and surface tension as suppression mechanisms for the classical KH instability.
In the limit of zero-net shear flow (flow oscillating around a mean of zero), the instability can be described by a Mathieu equation for the vertical displacement of the interface. This exhibits both a KH instability and a subharmonic parametric instability driven by a resonance between the surface gravity waves and the oscillating shear flow, for different parameters.

Parametric instabilities occur in many circumstances when there is periodic forcing in a system that supports waves. 
If the wave has small amplitude, this instability is caused by the triadic interaction of the primary wave with a pair of (typically) smaller-scale daughter waves.
For example, internal gravity waves in density stratified fluids, such as the Earth's oceans, are unstable to parametric instabilities that can transfer energy to smaller scales which are then dissipated \citep{McEwan1975,Drazin1977}. Another example is that of rotating fluids with elliptical streamlines, which can be unstable to the elliptical instability, a parametric instability involving the coupling of pairs of inertial waves with the elliptical deformation \citep{Kerswell2002}. Of most relevance to this paper is the parametric instability of Alfv\'{e}n waves. This instability is believed to play a role in reducing the correlation between velocity and magnetic field fluctuations in the solar wind as the waves propagate out into the heliosphere \citep[e.g.][]{MALARA1996}.
The decay of an Alfv\'{e}n wave via this instability has also been observed in experiments \citep{DORF2016}.

An important extension to the work of \citet{KELLEY1965} was performed by \citet{ROB1973}.
This work investigated the development of the parametric instability in an oscillating MHD flow where the flow direction is aligned with the magnetic field.
The magnetic field provides a tension that acts in a similar fashion to the surface tension treated in \citet{KELLEY1965}, working to suppress the KH instability and enabling the existence of surface Alfv\'{e}n waves. If these waves are resonant with the oscillation frequency of the shear flow, they become parametrically unstable.
\citet{ZARO2002} extended this concept to show that the parametric instability can drive the transfer of energy from fast magnetoacoustic waves into Alfv\'{e}n waves.
One possible application of the MHD parametric instability has been the investigation of periodic gravitational forcing resulting in a field-aligned flow. Parametric growth of oscillations was found to result in an enhanced strength of a magnetic field through the parametric instability, with application to the solar dynamo \citep{ZAQA2000,ZAQA2001,ZAQA2002}.

{In this paper we investigate how the presence of an oscillatory shear flow in the presence of a uniform magnetic field \emph{perpendicular} to the flow (i.e. the vorticity vector and the magnetic field vector are aligned)} can influence the development of the KH instability, or alternatively lead to parametric instabilities. We analyse the simplest model possible: a discontinuous oscillatory shear flow in a local Cartesian domain, and we derive the linear stability criteria analytically. We find that the oscillatory shear flow can be unstable to either the KH instability, or to a parametric instability involving the excitation of surface Alfv\'{e}n waves, depending on the parameters. Finally, we discuss the implications of our results for driving turbulence in the solar corona by kink waves.

\section{Model and Linear Stability Analysis}

\subsection{Model}

\begin{figure*}
  \begin{center}
\includegraphics[width=12cm]{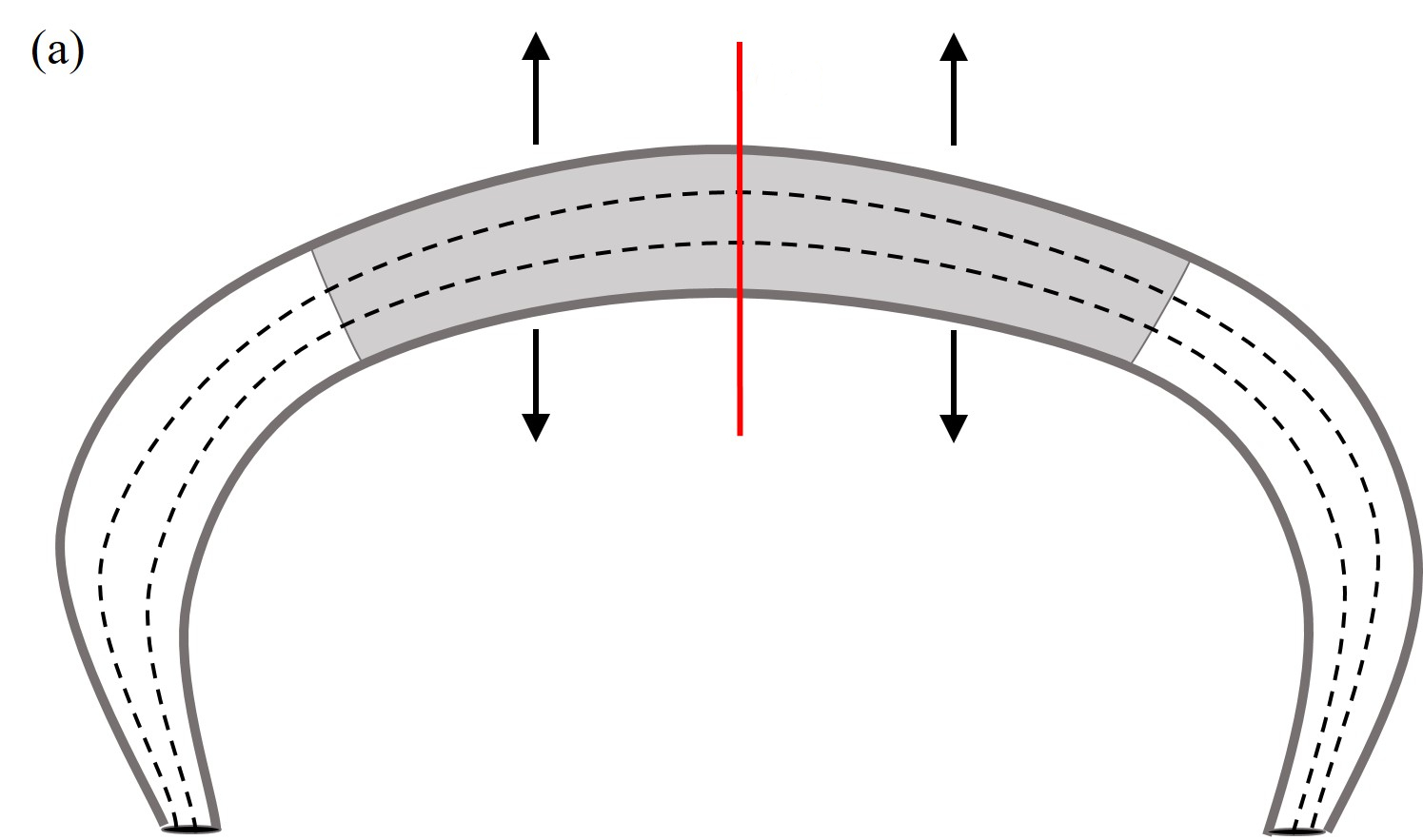}\\
\includegraphics[width=10cm]{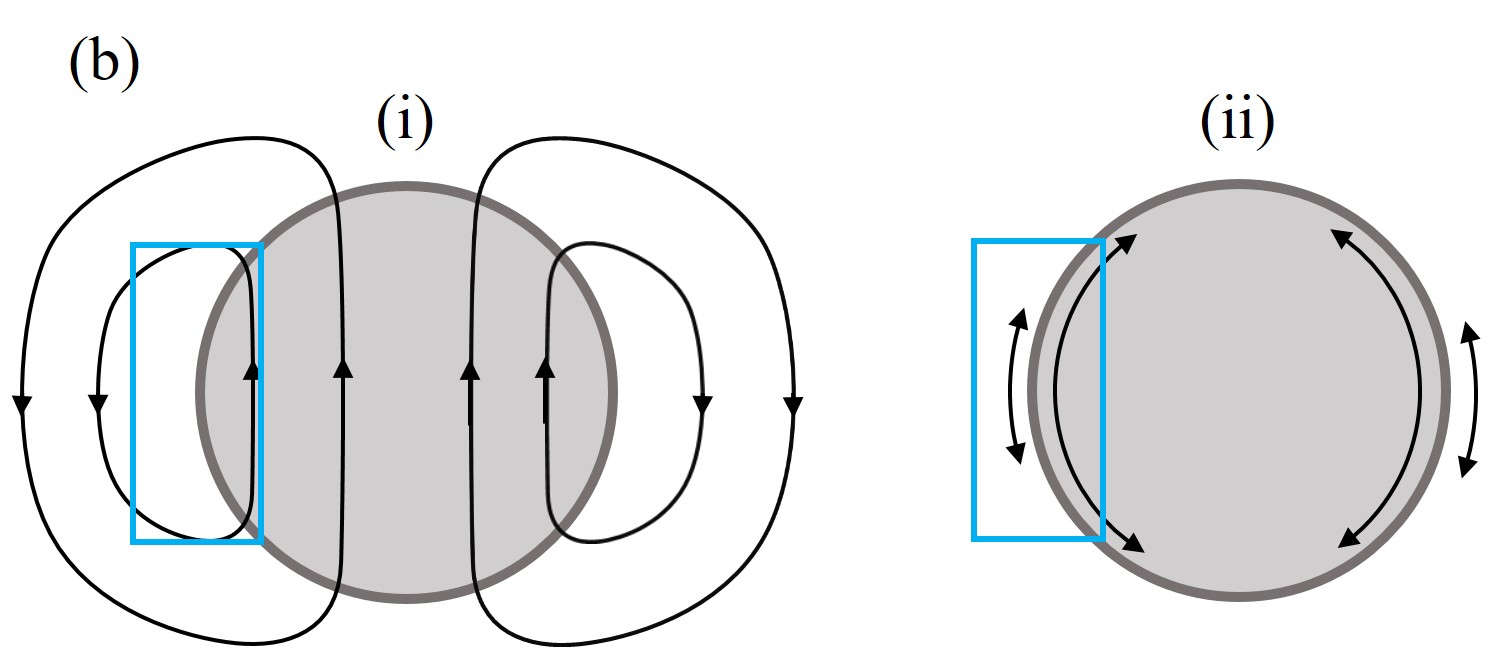}\ \ \ \ \ \ \ \ \ \ \ \ 
\includegraphics[width=4cm]{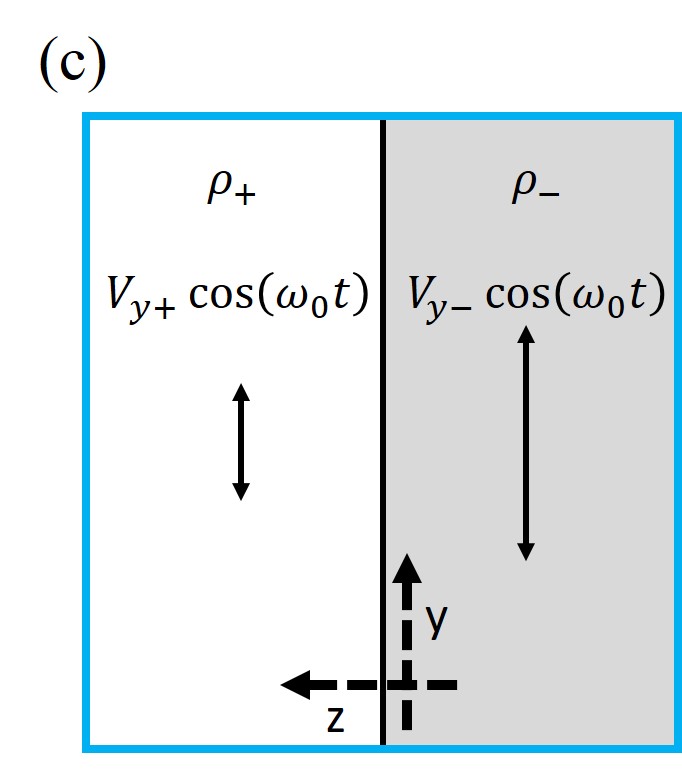}
  \end{center}
  \caption{Panel (a): a large scale schematic diagram of an oscillating flux tube rooted in the solar photosphere. The black dashed lines show the magnetic field lines and the black arrows show the direction of oscillation. The shaded region represents dense material in the tube. The red line marks the position of the cross-section displayed in panel (b). Panel (b): image of the cross-section of the flux tube with flow patterns. The lines and arrows show: (i) a snapshot in time of the streamlines of the dipole flow formed around a kink-oscillating flux-tube \citep[instability in this setting was investigated numerically by][]{TERR2008}, note that the direction of the flow arrows will reverse periodically with the wave motion, and (ii) the shear-flow in a flux tube associated with surface Alfv\'{e}n waves \citep[instability in this setting was investigated numerically by][]{ANTOLIN2015}. The blue boxes show the local region modelled in this paper as shown in panel (c). Panel (c): diagram of the local Cartesian model investigated in this paper with different densities and magnitudes of the velocity field in the regions above and below the density jump. Both regions have a velocity field that oscillates in phase at the same frequency and have the same magnetic field strength, both in the $x$-direction.} 
\label{setup}
\end{figure*}

Our motivation is to describe the development of the KH instability as this grows when driven by MHD oscillations in coronal flux tubes (an example of such a  situation is shown in Panel (a) of Fig.~\ref{setup}).
There are a number of possible flow profiles associated with oscillations of a flux tube, two are shown in the cross-sections shown in Panel (b) of Fig.~\ref{setup}. 
However, this is a complicated configuration which would be difficult to treat analytically.
In this paper, we investigate a simpler problem that provides a good approximation to the relevant dynamics of this system. We perform a local Cartesian analysis looking at the apex of the flux tube, where the amplitude of the velocity shear driven by the fundamental kink mode is largest, and the side of the flux tube with strong oscillatory shear flow with a setup shown in Fig.~(\ref{setup}) panel (c). Thus in this model we have an oscillatory shear flow in the presence of a uniform horizontal magnetic field. {Note that we neglect variations of the flow around and along the tube to allow us to make analytical progress. We also neglect the spatial and temporal variation of the magnetic field that would be associated with the magnetic oscillation. 
This is justified because the background flux tube is dominated by the axial component of the magnetic field.}

The next approximation we make is that the oscillations in our model are driven by a periodic force in the momentum equation, not
self-consistently via an MHD wave. As a result of our forcing there should be a pressure response in the system, which we neglect
because our focus is on incompressible disturbances (where the pressure response is small compared to the background pressure). This omission, however, could be problematic when considering velocities comparable with the fast-mode wave speed. We also neglect the magnetic field oscillations associated with the wave, as they are small in comparison to the background field along the flux tube, which should dominate
the dynamics.

The equations governing our model are
\begin{align}
\frac{\partial \rho}{\partial t} +\mathbf{V} \cdot \nabla \rho=&0,\\
\rho \frac{\partial \mathbf{V}}{\partial t}+\rho \mathbf{V}\cdot\nabla\mathbf{V}=&-\nabla P +\mathbf{J}\times\mathbf{B}+F(z,t)\mathbf{\hat{y}}, \label{mom}\\
\frac{\partial \mathbf{B}}{\partial t}=&\nabla \times (\mathbf{V} \times \mathbf{B}),\\
\nabla \cdot \mathbf{V}=&0,\\
\nabla \cdot \mathbf{B}=&0,
\end{align}
where $\mathbf{V}$ is the velocity field, $\mathbf{B}$ is the magnetic field, $\rho$ the density, $P$ the gas pressure, $\mathbf{J}$ the current density, and $F(z,t)\mathbf{\hat{y}}$ is a forcing term that maintains an oscillatory shear flow of frequency $\omega_0$. We consider a shear flow in Cartesian geometry with the flow in the $y$-direction, gradients in the flow in the $z$-direction, and the magnetic field in the $x$-direction.

We take a uniform density in each layer, with the density discontinuity aligned with that of the velocity, i.e. at $z=0$.
This is necessary for our basic oscillatory state to be an exact solution of the governing equations.
This state is described as follows:
\begin{align}
\rho_0=&\begin{cases} \rho_- & \mbox{if $z < 0$};\\ \rho_+ & \mbox{if $z > 0$}\end{cases}, \\
V_{y,0}=&\begin{cases} V_-\cos(\omega_0 t) & \mbox{if $z < 0$};\\ V_+\cos(\omega_0 t) & \mbox{if $z > 0$}\end{cases}, \\
V_{x,0}=&V_{z,0}=0,\\ 
P=&P_0,\\
B_{x,0}=&B,\\
B_{y,0}=&B_{z,0}=0.
\end{align}
We also define the velocity difference $\Delta V_0=V_+-V_-$. 
Here we note that the $y$-component of the background state equation resulting from Eq.~(\ref{mom}) is:
\begin{equation}
\rho_0 \frac{\partial {V_{y,0}}}{\partial t}=F(z,t).
\end{equation}
The other equations are trivially satisfied by the background state.


\subsection{Linear Analysis}

Linearising about the basic state of an oscillatory shear flow in the form $G=G_0+g$, where $G_0$ is a background state variable and $g$ is the corresponding linear perturbation, leads to the following set of equations:
\begin{align}
\frac{\partial \rho}{\partial t} +V_{y,0}\frac{\partial \rho}{\partial y}+v_z \frac{\partial \rho_0}{\partial z}=&0,\\
\rho_0 \frac{\partial \mathbf{v}}{\partial t}+\rho_0 V_{y,0}\frac{\partial \mathbf{v}}{\partial y}=&-\nabla p +\mathbf{j}\times\mathbf{B}_0,\\
\frac{\partial \mathbf{b}}{\partial t}=&-V_{y,0}\frac{\partial \mathbf{b}}{\partial y}+B\frac{\partial \mathbf{v}}{\partial x},\\
\nabla \cdot \mathbf{v}=&0,\\
\nabla \cdot \mathbf{b}=&0.
\end{align}

{Taking perturbations to be normal modes of the form $f(x,y,z,t)=\tilde{f}(z,t)\exp(ik_xx+ik_yy)$ for the scalar quantities and $\mathbf{f}(x,y,z,t)=\tilde{\mathbf{f}}(z,t)\exp(ik_xx+ik_yy)$
for vector quantities, we may derive an equation for the temporal evolution of the vertical Lagrangian displacement of the fluid ($\tilde{\eta}_z$) which relates to the z-component of the velocity $\tilde{v}_z=\partial\tilde{\eta}_z/\partial t$.
Using that the physical variables are constant in the regions both above and below the discontinuity and the requirement that the perturbation decays to $0$ at $z=\pm \infty$ in ideal MHD gives the $z$ dependence of the eigenfunction as $\exp(-k|z|)$.
Therefore we can define $\tilde{\eta}_z(z,t)=\eta(t)\exp(-k|z|)$, and by matching the solutions across the interface, the following equation for $\eta$ is found:}
\begin{align}
\frac{d^2 \eta}{d t^2}+&2ik_y(\alpha_+V_++\alpha_-V_-)\frac{d \eta}{d t} +\left[ik_y\left(\alpha_+\frac{d V_+}{d t}+\alpha_-\frac{d V_-}{d t} \right) \right.\\-&\left.k_y^2(\alpha_+V_+^2+\alpha_-V_-^2) +\frac{k_x^2B^2}{2\pi (\rho_++\rho_-)}\right]\eta=0,\nonumber
\end{align}
where $\alpha_{\pm}=\rho_{\pm}/(\rho_++\rho_-)$. The derivation of this equation is presented in Appendix \ref{lin_deriv}.
If we set $\eta=\hat{\eta}\exp(-ik_y\int\alpha_+V_++\alpha_-V_-dt)$, upon rearranging we get:
\begin{equation}\label{Mathieu}
\frac{d^2 \hat{\eta}}{d t^2} +\left[\frac{k_x^2B^2}{2\pi (\rho_++\rho_-)} -\frac{1}{2}k_y^2\alpha_+\alpha_-\Delta V_0^2\left(1+\cos (2 \omega_0 t)\right) \right]\hat{\eta}=0. 
\end{equation}

{Using the Alfv\'{e}n speed in the $+$ region, i.e. $V_{\rm A+}=\sqrt{B^2/4\pi\rho_+}$, and the wavenumber $k_0$ we can determine the following dimensionless quantities:
\begin{align}
t=&\frac{T}{V_{\rm A+}k_0},\\
\Delta V=&V_{\rm A+}M_{\rm A},\\
k_x=&k_0K_x,\\
k_y=&k_0K_y,\\
\hat{\eta}=&\frac{\eta'}{k_0},
\end{align}
where $T$, $\Delta V$, $K_x$, $K_y$, and $\eta'$ are dimensionless and $M_{\rm A}=\Delta V/V_{\rm A+}$ is the Alfv\'{e}nic Mach number.
For cases where $\omega_0\ne0$ we are free to select $k_0$ such that $\omega_0=1$, giving the dimensionless equation:}
\begin{equation}\label{Mathieu_norm}
\frac{d^2 {\eta}'}{d T^2} +\frac{\alpha_+}{2}\left[4K_x^2 -\alpha_-K_y^2M_{\rm A}^2\left(1+\cos (2 T)\right) \right]{\eta}'=0. 
\end{equation}
This is a Mathieu equation, which takes the general form:
\begin{equation}\label{standard_mat}
\frac{d^2f}{dT^2}+(a-2\varepsilon\cos(2T))f=0,
\end{equation}
where $a$ and $\varepsilon$ are constants.
Eq.~(\ref{Mathieu_norm}) is equivalent to Eq.~(\ref{standard_mat}) if $f=\eta^{\prime}$, and:
\begin{align}
a=&2\alpha_+K_x^2 -\frac{1}{2}\alpha_+\alpha_-K_y^2M_{\rm A}^2,\label{a_defn}\\
\varepsilon=&\frac{1}{4}\alpha_+\alpha_-K_y^2M_{\rm A}^2.\label{epsilon_defn}
\end{align}
This realisation is useful because the properties of the Mathieu equation are well understood \citep[e.g.][]{BENDER1978}.

\subsection{General solutions to the Mathieu Equation}\label{Mat_anal}

To understand the linear instabilities of an oscillating shear flow, it is helpful to first consider the general Mathieu equation (Eq.~(\ref{standard_mat})), and its resulting instabilities.
We can rewrite Eq.~(\ref{standard_mat}) as:
\begin{equation}\label{mod_gen_Mat}
\frac{d^2f}{dT^2}+af=2\varepsilon\cos(2T)f=\varepsilon\left(\exp(2iT)+\exp(-2iT) \right)f.
\end{equation}
The solutions to this equation must obey Floquet's theorem, i.e.~ $f=C_1\exp(i\omega T)\phi(T)+C_2\exp(-i\omega T)\phi^*(T)$, where $\phi(T)$ is a function that is periodic with the same periodicity as the time-varying coefficients, $C_1$ and $C_2$ are arbitrary constants, and $*$ denotes the complex conjugate. Here $\phi(T)$ is given by $\phi(T)=\Sigma_{p=-\infty}^{\infty}A_p\exp(2\mathrm{i}p T)$, where $p$ is an integer.
Using this solution to $f$, inductive solutions to Eq.~(\ref{mod_gen_Mat}) can be determined:
\begin{equation}\label{induc_soln}
\left[-(\omega +2p)^2+a\right]A_p=\varepsilon(A_{p-1}+A_{p+1}).
\end{equation}
{This is an infinite system of equations for the coefficients $A_p$ for each $p$. One can determine $\omega$ from the requirement that this system has non-trivial solutions. These consist of tongues of instability centred on certain frequencies.}

{Analytic solutions to these equations can be obtained in the limit $\varepsilon \ll 1$ from Eq.~(\ref{induc_soln}). For $p=0$, we must have:
\begin{equation}
\omega^2\approx a,
\end{equation}
which describes an oscillation with $\omega=\pm \sqrt{a}$.
For $p\ne 0$ this implies that to have a non-zero coefficient $A_p$, we must have:
\begin{equation}
a=p^2,
\end{equation}
i.e. there is a resonance between the excited wave and the oscillatory forcing, which is a parametric instability. More generally, it can be shown that instability occurs within fingers of instability (for $a>0$) that are bounded by
\begin{equation}
a=p^2 \pm \varepsilon + O(\varepsilon^2),
\end{equation}
and the maximum growth rate at the centre of the dominant $p=1$ resonance is
\begin{equation}
\sigma\equiv \mathrm{Im}[\omega]=\frac{\varepsilon}{2} + O(\varepsilon^2),
\end{equation}}
\citep[see, for example,][]{BENDER1978}.

\section{Exploring the nature of the instabilities}

Now that we have an ODE in the form of a Mathieu equation, we can explore the consequences of having an oscillatory shear flow. To gain intuition, it is helpful to consider the case of constant shear (i.e. setting $\omega_0=0$ so that the term $\cos 2\omega_0t =1$ in Eq.~(\ref{Mathieu})).
In this regime we select our normalising wavenumber $k_0$ to be an arbitrary real wavenumber greater than 0.
In this case the dimensionless Mathieu equation becomes:
\begin{equation}
\frac{d^2 {\eta}'}{d T^2} +\alpha_+\left[2{K_x^2} -\alpha_-K_y^2M_{\rm A}^2 \right]{\eta}'=0,
\end{equation}
which has constant coefficients and normal mode solutions of the form 
${\eta'}\propto\exp(\pm i\omega T)$. This leads to:
\begin{equation}\label{term_check}
\omega^2={\alpha_+}\left(2K_x^2-\alpha_-K_y^2M_{\rm A}^2\right),
\end{equation}
where the first term arises from magnetic tension and is stabilising, while the second comes from the shear and is destabilising.
In this equation the first term on the RHS is the square of the MHD kink wave frequency in the incompressible limit, which describes a surface Alfv\'{e}n wave, and the second term describes the Doppler-shifting of this wave by the shear flow.
When the second term becomes larger than the first, $\omega^2$ is negative and the system is unstable to the MHD KH instability with a growth rate given by $|\omega|$ \citep[e.g.][]{CHAN1961}.
This can be mathematically stated as the following condition for the onset of instability:
\begin{equation}\label{KH_limit}
M_{\rm A}^2>\frac{2K_x^2}{\alpha_-K_y^2}.
\end{equation}
Simply put, if the Alfv\'{e}nic Mach number becomes sufficiently large, for a given wavevector direction, then the shear flow can overcome magnetic tension and the system becomes unstable.
If there is no perturbation at all in the direction of the magnetic field ($K_x=0$) then there is no suppression by magnetic tension and the system is unstable for \emph{any} non-zero velocity difference. But as the angle of the wavevector to the magnetic field tends towards zero the driving force is reduced and the tension force is increased so the system tends towards stability.
The $K_x=0$ modes are rather special as they would correspond in our model to modes with no, or possibly global, variation along the flux tube; instead, their variation is confined primarily to around the circumference of the flux tube.

\begin{figure*}
  \begin{center}
\includegraphics[width=17cm]{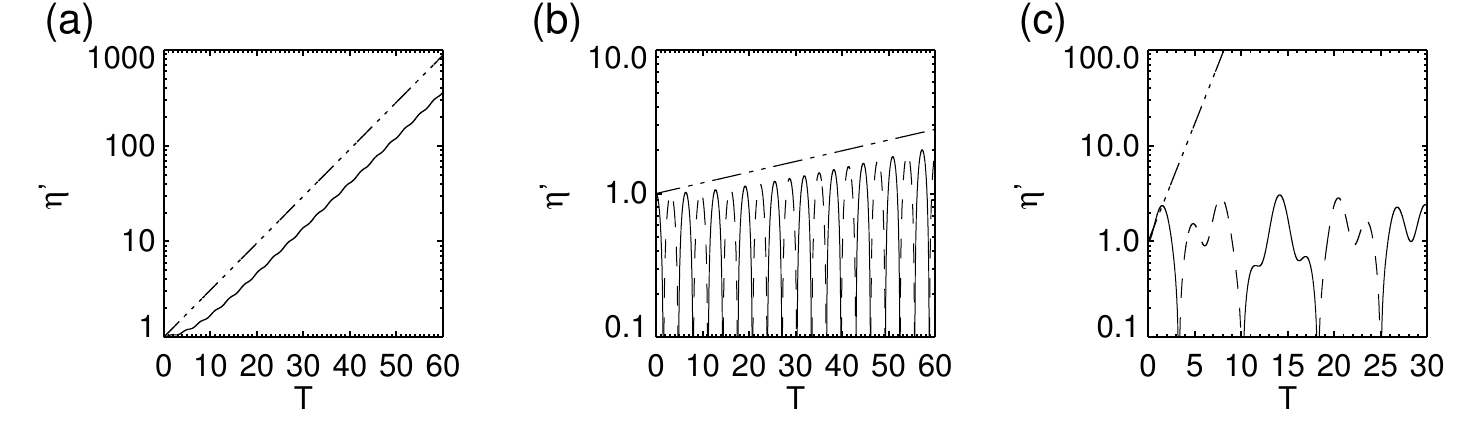}\\
\includegraphics[width=17cm]{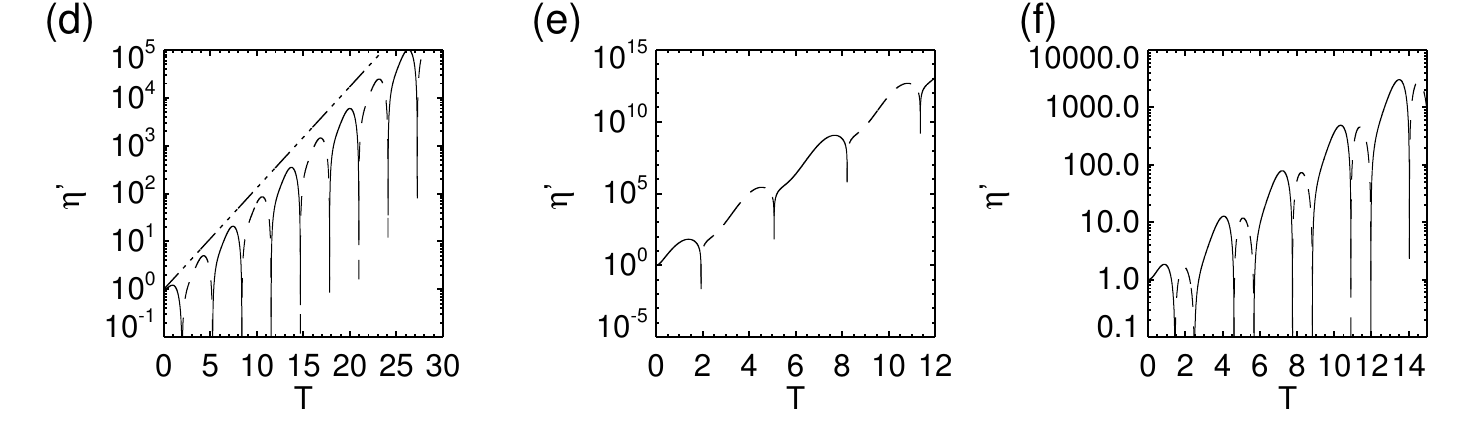}
  \end{center}
  \caption{Calculation of the variation of $\eta'$ with $T$ for $M_{\rm A}=0.2$, and $\alpha_+=0.01$ using $\theta=1.48$ (where $\theta=\tan^{-1}(K_y/K_x)$) in panels (a) and (c), $\theta=1.20$ in panel (b), $\theta=1.45$ in panel (d), $\theta=1.5$ in panel (e), and $\theta=1.45$ in panel (f). $K=20$ (where $K=\sqrt{K_x^2+K_y^2}$) is used for panels (a) and (b), $K=100$ for (c) and (d), $K=300$ for (e) and $K=200$ for (f). The dash-triple dot line in panels (a) to (d) gives the predicted growth rate. The solid lines show positive and the dashed lines negative values of $\eta'$. The positions in $K$-space for each of these panels are marked on Fig.~(\ref{instability}).}
\label{calc_growth_rate}
\end{figure*}

{When the oscillatory term is included (i.e. $\omega_0\ne 0$, and we solve Equation~\ref{Mathieu_norm}), then as in \citet{KELLEY1965} and \citet{ROB1973} then both the KH instability and parametric instabilities will be possible.
This can be seen in Fig.~(\ref{calc_growth_rate}), which gives numerical solutions to the temporal evolution of ${\eta}'$ from solving Eq.~(\ref{Mathieu_norm}) first by splitting this equation into two coupled equations for $\eta'$ and $d \eta'/d T$ and solving these with a first order, forward difference solver.
Solutions are given for $M_{\rm A}=0.2$, and $\alpha_+=0.01$ for different points in $\mathbf{K}$-space (these points in $\mathbf{K}$-space are shown in Fig.~\ref{instability}).
These show a KH unstable mode (panel a), $p=1$ parametric unstable modes (panels b, d, e), $p=2$ parametric unstable mode (panel f), and a stable mode (panel c).
Looking at this figure, it is clear to see that the KHI is a direct instability of our system (shown by the fact there is only a solid black line in panel a), but the parametric instability involves a resonantly enhanced wave so the solution takes both positive and negative values (see the switch between solid - positive $\eta'$ - and dashed - $-\eta'$ - lines in panel b). 
We explore the different instability behaviour in this section.}

\subsection{Properties of the instabilities in the limit of weak shear}

In this section we consider a limit of Eq.~(\ref{Mathieu}) where the amplitude of the oscillatory shear flow (proportional to $\cos 2 T$) is small in a similar fashion to Section \ref{Mat_anal}, as also considered by \citet{KELLEY1965} and \citet{ROB1973}. We use this to develop solutions for the growth rate of the two instabilities. Strictly speaking, we take $\varepsilon \ll 1$ (as defined in Eq.~(\ref{epsilon_defn})) and follow the method presented in \citet{BENDER1978}. Physically this corresponds to the limit where the square of the shear rate is small compared to the square of the oscillation frequency, and one example where this may occur is when the wave driving the shear flow has a small amplitude.

\subsubsection{Magnetic KH instability}\label{KH_mode}

From Section \ref{Mat_anal} when the resonance condition is not satisfied then the wave frequency is given by $\omega=\pm\sqrt{a}$ and by direct comparison it is possible to state the frequency of the surface wave in the small shear limit.
This is given as:
\begin{equation}\label{wave_KH}
\omega=\pm\sqrt{\frac{\alpha_+}{2}}\sqrt{4K_x^2-\alpha_-K_y^2M_{\rm A}^2}.
\end{equation}
This is very similar to Equation \ref{term_check}, the only difference being that the second term underneath the square root is a factor of $1/2$ smaller, that results from averaging the shear kinetic energy over one period. 

To have a direct instability of the system, the stiffness of the boundary between the two flows (i.e. the flux tube boundary) needs to disappear, i.e.:
\begin{equation}
M_{\rm A}^2>\frac{4K_x^2}{\alpha_-K_y^2},
\end{equation}
which means that the non-oscillatory term in Eq.~(\ref{Mathieu}) is negative.
For instability the Alfv\'{e}nic Mach number has to be $\sqrt{2}$ greater than the case with the same wavevector $\mathbf{K}$ for a non-oscillatory shear flow (see Eq.~(\ref{KH_limit})).
As can be expected for the MHD KH instability, the shorter the wavelength in the direction of the flow and the longer the wavelength in the direction of the magnetic field the more likely the system is to be unstable.
Though it is important to remember that for this inequality to accurately describe the onset of direct instability, $\frac{1}{4}\alpha_+\alpha_-K_y^2M_{\rm A}^2 \ll 1$.
Figure~(\ref{calc_growth_rate}) panel (a) compares the predicted growth rate of the KH instability to the solution of Eq.~(\ref{Mathieu_norm}) showing that for these parameters the solution in the asymptotic limit and the solution match well.
For panel (c), though predicted to be unstable in the asymptotic limit, the KH mode does not grow for these parameters as they are beyond the applicability of this limit. We will look at this further in Section \ref{Adrian}.
Note that $\theta=\tan^{-1}(K_y/K_x)$.

\subsubsection{Subharmonic resonance}\label{para_mode}

If the system is stable to the KH instability, then a perturbation takes the form of a surface wave instead.
{The frequency of the surface shear Alfv\'{e}n wave is given by Eq.~(\ref{wave_KH}).
There exists a parametric resonance between this wave and the oscillatory driver when, as stated in Section \ref{Mat_anal}, the following condition is satisfied:
\begin{equation}\label{para_cond}
\omega^2=a=p^2, \ \ \mbox{for} \ \ p=1, 2, 3, ...
\end{equation}
Note this is different to the resonance process behind resonant absorption discussed in Section \ref{INTRO}.
As the strongest resonance occurs for $p=1$, then the fastest growing modes satisfy:
\begin{equation}
\omega= \pm1,
\end{equation} 
which is equal to half the frequency of the oscillatory forcing, indicating subharmonic resonance.
This corresponds to counter-propagating surface Alfv\'{e}n waves that become unstable if their frequency magnitudes are equal to the frequency of the oscillatory shear flow.}

The parametric instability is not just confined to the exact resonance, there is an envelope around this exact resonance that is unstable (see e.g. Section 2.3), with the width of the envelope being given by $2\varepsilon$ \citep[e.g.][]{BENDER1978}.
We note here that for a larger Alfv\'{e}nic Mach number or $\alpha_+\sim\alpha_-$ then more of the parameter space is unstable to parametric instabilities because $\varepsilon$ becomes larger.
From this we can define the following band where the dominant subharmonic parametric instability is possible:
\begin{equation}
\frac{3}{8}\alpha_-K_y^2M_{\rm A}^2>K_x^2-\frac{1}{2\alpha_+} >\frac{1}{8}\alpha_-K_y^2M_{\rm A}^2.
\end{equation}

The maximum growth rate of the instability happens when the exact resonance condition is satisfied.
This growth rate is given by \citep[e.g.][]{BENDER1978}:
\begin{equation}\label{max_parametric}
\sigma_{\rm max}=\frac{\varepsilon}{2}=\frac{1}{8}\alpha_+\alpha_-K_y^2M_{\rm A}^2.
\end{equation}
Of note here is the dependence of the growth rate on $M_{\rm A}^2$. 
This is contrary to the KH modes, which grow at a rate proportional to $M_{\rm A}$.
Figure~(\ref{calc_growth_rate}) panels (b) and (d) compare the predicted growth rate of the parametric instability to the solution of Eq.~(\ref{Mathieu_norm}) with a first order forward difference solver for the same parameters with the solution in the asymptotic limit proving to be accurate.

From this analysis we can see that the parametric instability is quite different to the magnetic KH instability. While the latter grows fastest for modes that minimise $K_x$ (i.e. wavelengths along the magnetic field as long as possible are preferred), modes unstable to the parametric instability must have finite $K_x$. Therefore it is the parametrically unstable modes that are good at kinking and disturbing the boundary between the two flows \emph{along} the direction of the magnetic field.

\subsection{Solutions to the Mathieu Equation}\label{Adrian}

\begin{figure*}
  \begin{center}
\includegraphics[width=13cm,trim={4.5cm 0 4cm 0}, clip]{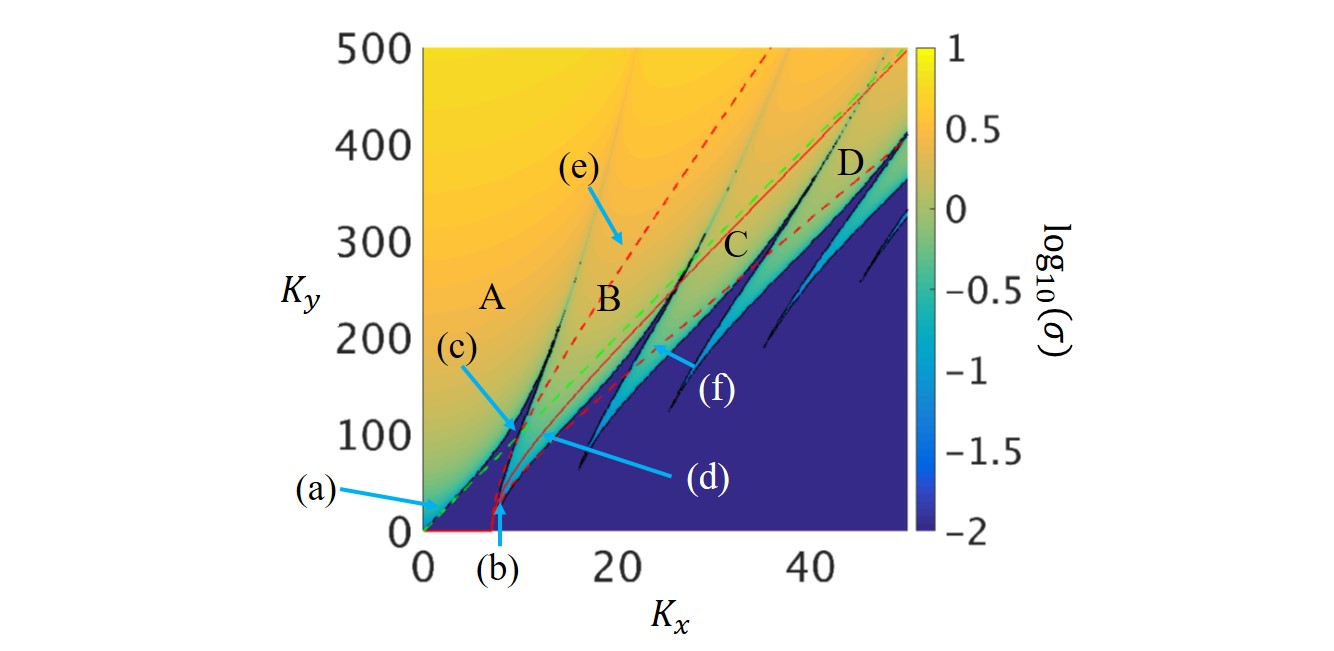}
  \end{center}
  \caption{Numerically computed logarithm of the growth rate in $K$-space. Blue regions are approximately stable (with $\sigma\le10^{-2}$). The top left region {(which emanates from $K_x=K_y=0$)} is KH unstable, whereas the five fingers of instability (yellowish regions pointing down) to the right of the KH band  (associated with $K_x=p/\sqrt{2\alpha_+}$ for $K_y=0$) are parametrically unstable. These fingers of instability are associated with $p=1$ to $5$ in equation \ref{para_cond}.
The capital letters A to D mark different instability bands, where A= KH instability, B= Parametric (p=1), C= Parametric (p=2), D= Parametric (p=3). The green dashed line marks the critical wavenumber for the growth of the KH instability in the asymptotic limit. The solid red line gives the fastest growing parametric mode in the asymptotic limit, with the dashed red lines marking the band of approximate resonance valid for small $K_y$. In this calculation $\alpha_+=0.01$ and $M_{\rm A}=0.2$. The labelled blue arrows show the position in $K$-space of the six calculations shown in Figure \ref{calc_growth_rate}.}
\label{instability}
\end{figure*}

Figure~(\ref{instability}) plots the base 10 logarithm of the growth rate and shows the regions of instability on the $(K_x,K_y)$-plane. 
The values were computed by solving Eq.~(\ref{Mathieu_norm}).
In this figure, $\alpha_+=0.01$ and $M_{\rm A}=0.2$. 
We analyse Eq.~(\ref{Mathieu_norm}) numerically using a Floquet method. First, we write Eq.~(\ref{Mathieu_norm}) as a system of 2 coupled ODEs. 
We then compute the monodromy matrix of linearly independent solutions, which is accomplished by integrating the ODEs over one period $\pi$ for initial conditions such that all variables except one are set to zero (using a 4/5th order Runge--Kutta method). The eigenvalues of the monodromy matrix allow us to obtain the complex growth rates of the instability.
As can be seen in Fig.~(\ref{instability}) there are regions of stability (blue) but also bands of instability shown in green and yellow.

In Fig.~(\ref{instability}) there are a number of bands in $K$-space where the system becomes unstable.
There are two distinct instabilities that exist in the unstable bands.
In the band in the top left (associated with small $K_x$), above the green dashed line, the system is KH unstable (see panel (a) of Fig.~(\ref{calc_growth_rate})).
However, the other bands are related to the resonant growth of waves (see panels (b) and (d) of Fig.~(\ref{calc_growth_rate})).
Panel (c) shows a region that is expected to be KH unstable based on the asymptotic limit, but the instability is switched off as it is impinged by the resonance.
The instability bands are well separated for small $K_y$, but as $K_y$ increases the resonance impinges on the KH instability.
In Fig. \ref{instability} the dashed green line marks the theoretically predicted cutoff for the KH instability (see Section \ref{KH_mode}) and the solid red line with the dashed red lines mark the $p=1$ subharmonic resonance growing through a parametric instability (see Section \ref{para_mode}).
For the parameters used in the calculation of Fig. \ref{instability} the asymptotic limits are expected to hold when $K_y \ll 100$ (\textbf{i.e.~$\varepsilon\ll1$}).


From Fig.~(\ref{instability}) we can see that, though crude, where the upper bound of the parametric instability as calculated by the asymptotic limit meets the marginal stability curve of the KH instability (again in the asymptotic limit) can be used as an approximate wavevector for where the resonance begins to impinge on the KH instability.
This wavevector is given by:
\begin{align}
K_x=&\frac{1}{\sqrt{\alpha_+}},\\
K_y=&\frac{2}{\sqrt{\alpha_{+}\alpha_{-}}}\frac{1}{M_{\rm A}}.
\end{align}

{Figure~\ref{instability} shows a total of five parametrically unstable regions, which are associated with $p=1$ to $5$ in Eq.~(\ref{para_cond}).
Each of these fingers of instability start from $K_y=0$, i.e. from the $x$-axis, but because as $p$ gets larger the resonance gets weaker the fingers of instability get thinner. This makes them harder to accurately plot in the figure near $K_y=0$ as they are more challenging to resolve in $K$-space.
For a given $K_y$, we find that the largest growth rate is associated with a KH mode, with the $p=1$ resonance next largest, and the growth rate getting smaller as $p$ increases.}

{In the region of the figure above the dashed green line, the asymptotic limits predict that this is unstable everywhere, and this is generally correct. However, it is important to note that even in this regime, where $a$, as defined in Eq.~(\ref{a_defn}), is negative, the regions of parametric instability appear.
Even with $a<0$ these parametric unstable regions still represent the exponential growth of an oscillation around zero and not a purely growing mode (as shown in Fig.~(\ref{calc_growth_rate}) panel (e)). Compared to the non-oscillatory case, the key change in stability of the system comes when $a$ is positive, i.e. below the dashed green line, and resonances can drive perturbations to the system to grow.}

\section{Application to the Solar Atmosphere}

Now let us take the model back to its application to coronal flux tubes. When considering the instability on an oscillating flux tube in the solar atmosphere, firstly the different scales of the problem need to be considered.
For a prominence thread, the diameter ($D$) is often of the order of $2\times 10^2$\,km \citep[e.g.][]{ARREGUI2018} but can be up to $10^3$\,km \citep[e.g.][]{OKA2015}.
The length ($L$, taken as the distance of the field lines leaving the photosphere to their return) can be two or more orders of magnitude greater than $D$. 
The length of the flux tube filled with dense material is often found to be between $\sim 3\times10^3$ and $3\times 10^4$\,km \citep{ARREGUI2018}.
If the wavenumber associated with the instability on the flux-tube surface is considered, then it must necessarily have a larger aspect ratio with the flux-tube length than the diameter.
Roughly speaking these numbers hold for coronal loops (flux tubes that are filled with hot material) as well.


{To use this aspect ratio to constrain our calculations, we need to determine a lower limit for the normalised wavenumber $K$ given by the length of the flux tube in normalised form.
If a flux-tube is oscillating in the solar corona, the characteristic frequency of the oscillation is the kink frequency and the characteristic speed is the kink speed.
As the non-dimentional kink speed of a flux tube is $\sqrt{2\alpha_+}$, the frequency of the surface Alfv\'{e}n wave will, because it is driven by resonance with the kink-wave, equal the frequency of a global linear kink wave of the flux tube and have a normalised value of unity.
Therefore, the $K_x$ value (for $K_y=0$) that gives this is $K_{x,{\rm KINK}}=1/\sqrt{2\alpha_+}$, which is the $K_x$ value associated with the first resonance for $K_y=0$.
By taking $K_x\ge K_{x,{\rm KINK}}$ then we can look at wavenumbers that are appropriate for comparing to instabilities in coronal flux tubes.}
{For an aspect ratio of the flux tube of $1:100$ then instability modes must have $K_y\ge K_{y, {\rm D}}=1/D=200/\sqrt{2\alpha_+}$ where $D$ is the diameter of the flux tube.}

\begin{figure*}
  \begin{center}
\includegraphics[width=8.6cm,trim={4cm 0 4.5cm 0}, clip]{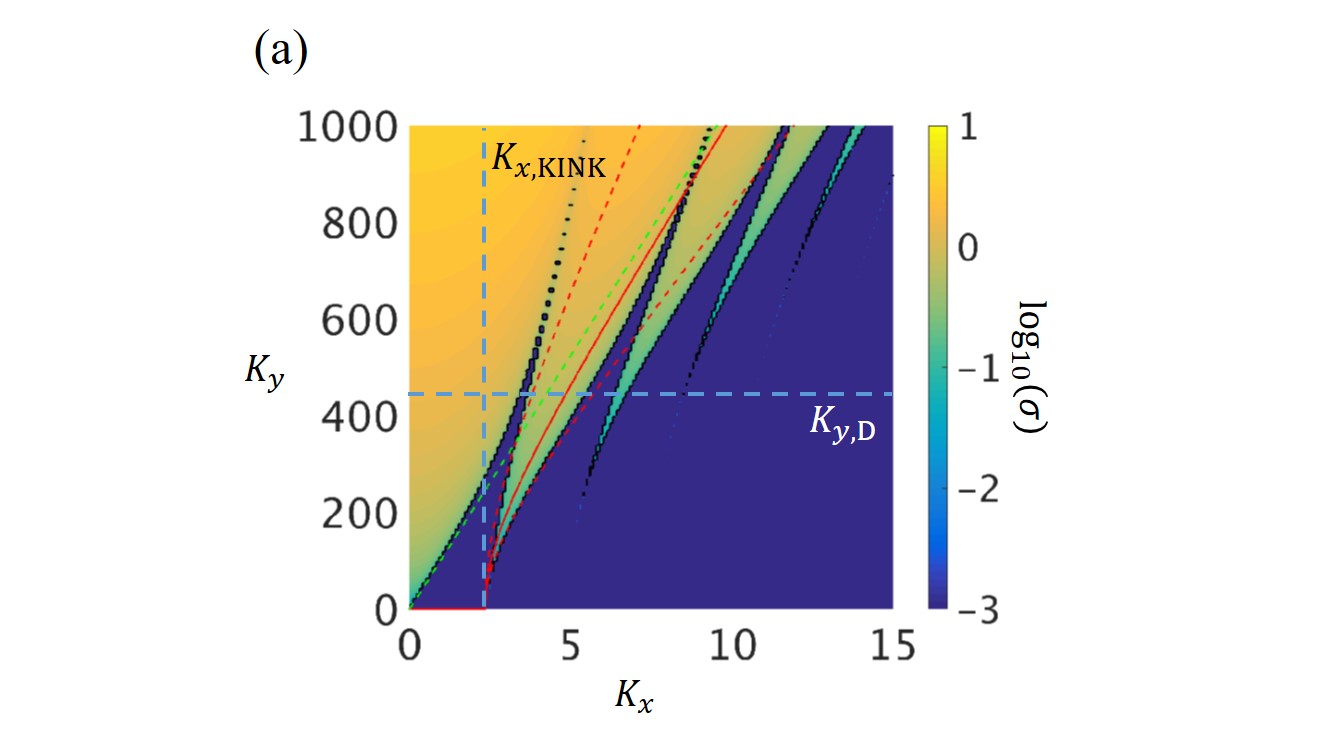}\ \ \ \
\includegraphics[width=8.6cm,trim={4cm 0 4.5cm 0}, clip]{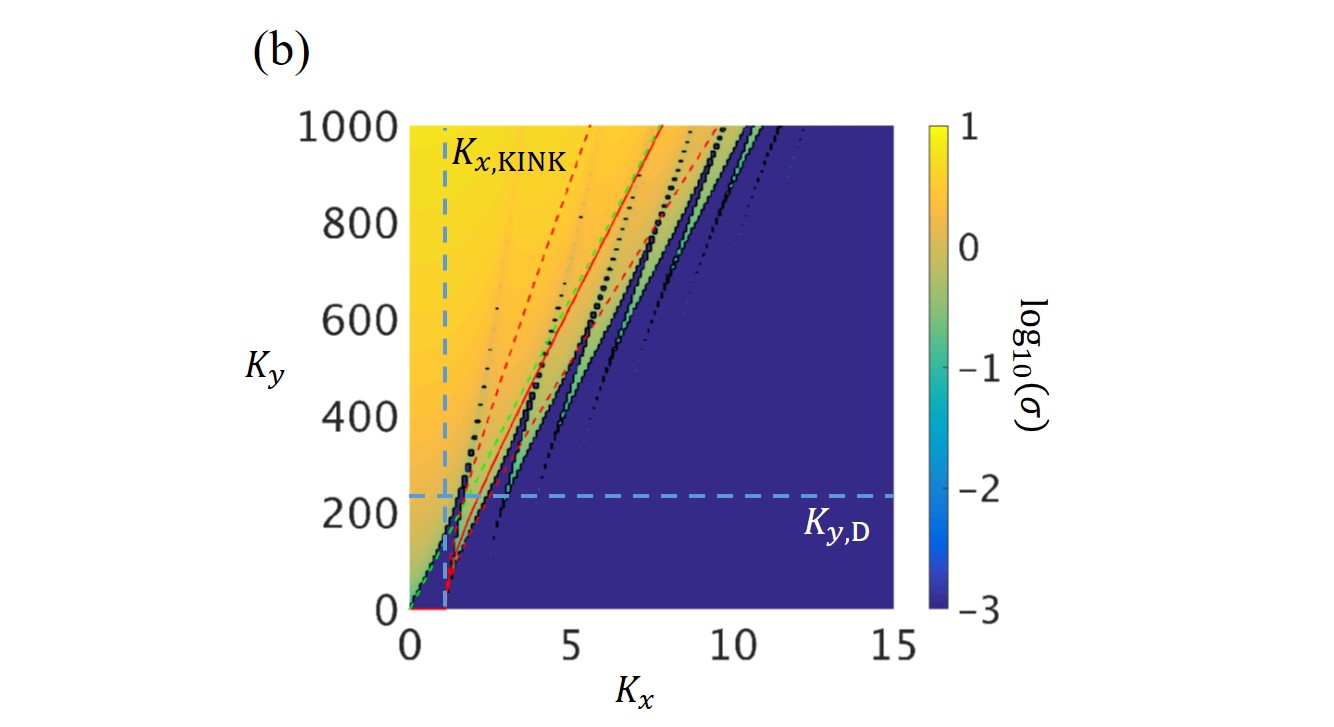}
  \end{center}
  \caption{Numerically computed logarithm of the growth rate in $K$ space for (a) $\alpha_+=1/11$ and $M_{\rm A}=0.02$ and (b) $\alpha_+=2/5$ and $M_{\rm A}=0.02$. In each panel the green dashed line marks the critical wavenumber for the growth of the KH instability in the asymptotic limit. The solid red line give the fastest growing parametric mode in the asymptotic limit, with the dashed red lines marking the band of approximate resonance. The vertical dashed blue lines show the position of $K_{x,{\rm KINK}}$ and the horizontal dashed blue lines show the position of $K_{y, {\rm D}}$. The top right quadrant, which is most relevant for instability in a solar setting, is dominated by parametric modes.}
\label{instability_obs}
\end{figure*}

Figure~(\ref{instability_obs}) shows calculations with density ratios of 1:10 (panel a) and 2:3 (panel b), and $M_{\rm A}=0.02$, calculated using a coronal Alfv\'{e}n speed of $1000$\,km\,s$^{-1}$ and a velocity difference of $20$\,km\,s$^{-1}$.
Generally the structure of the bands of instability is very similar to that shown in Fig.~(\ref{instability}).
Both $K_{x,{\rm KINK}}$ and $K_{y, {\rm D}}$ are marked with dashed blue lines, and we expect that instabilities on a flux tube will have to be in the quadrant that is to the right of the dashed blue line marked $K_{x,{\rm KINK}}$ and above the dashed blue line marked $K_{y, {\rm D}}$. 

If we only look at modes that use $K_{x,{\rm KINK}}$ then for all $K_y$ there is no parametric instability only KH instability. This instability provides the largest growth rate for a given $K_y$, but for all $K_y$ the growth rate is reduced compared to Eq.~(\ref{wave_KH}) as a result of the presence of the resonance, i.e. for all achievable growth rates of the KH instability on a flux-tube, the oscillatory nature of the shear and the impinging parametric resonance band reduces the growth of the KH instability.
These large scale modes are strictly outside the domain of applicability of the local model we have developed. To fully understand the dynamics of this global mode then the boundary conditions and variations along the flux tube would have to be included.

If we look at smaller scales along the flux tube, which is more consistent with the local model we employ, this would mean that we should use $nK_{x,{\rm KINK}}$ as the wavenumber along the magnetic field where $n$ is a integer greater than $1$. This puts the system into a regime where parametric instabilities dominate at the smallest unstable $K$ values (which, due to the larger spatial scales involved, can extract more energy from the shear flow).
Because these instabilities exist for different wave vectors, it can be expected that both of the instabilities could be growing in a system at the same time.

{For an instability, either KH or parametric, with $2K_{y, {\rm D}}\le K_y\le 4K_{y, {\rm D}}$, from Fig.~(\ref{instability_obs}) the growth rate is in the range $\sigma =|i\omega|\sim 1$ to $10\times \omega_0$.
Taking a characteristic oscillation period of a kink wave in the corona to be $300$\,s, then the time scale for the instability is approximately $30$ to $300$\,s.}
Note that these parametric modes are associated with positions in $K$-space above the dashed green lines in Fig.~(\ref{instability_obs}). For perturbations below that line (i.e. regions that would be stable for a non-oscillatory shear flow) the growth rate will be $\sigma\le \omega_0$, and so these perturbations can grow (with high wave number along the magnetic field) at time scales longer than $300$\,s and as such can still occur on dynamically important times scales.
Prominence threads may have a density up to $100$ times greater than the solar corona \citep[e.g.][]{PARENTI2014, ARREGUI2018}, for this case the structure of the instability bands would not change significantly and we would expect instability to grow on a time scale of approximately $30$\,s, as estimated using Fig. \ref{instability}.  
As this time scale will connect to the rotation time of a vortex at that scale {(i.e. the eddy turnover time)}, this time scale can also be used as a lower estimate of the turbulence time scale (and with it the heating time scale) of the flux tube, though nonlinear analysis is necessary to accurately estimate this value.

It is important to note here that for instability in a prominence thread, we have imagined a loop in the solar corona comprised of a flux tube that is completely filled by dense material, and used this to connect the length scales of the first resonance in the model to the loop length (or thread length as they are treated as being the same).
However, the reality is that only a section of the loop will contain this material \citep[e.g.][]{ARREGUI2018}.
\citet{SOLER2010b} investigated the change in period of a kink-wave as a result of the prominence thread only filling part of the flux tube, finding that the frequency of the wave changes.
The reduction in the period of a fundamental kink mode by a factor of two if the dense thread only fills 10 per cent of the flux tube instead of the whole tube \citep{ARREGUI2011}. This change in frequency from the local kink frequency will mean that the position of the resonance relative to $K_{x.{\rm KINK}}$ will change, but as this is not a large change it would not be expected to greatly influence our estimates.

\section{Summary and Discussion}

This paper demonstrates that an oscillating MHD shear flow is unstable to not only the KH instability, but also to parametric instabilities involving surface Alfv\'{e}n waves. 
In general the growth rate of the KH instability is larger than that of the parametric instability, but as $\varepsilon$ (which quantifies the magnitude of the shear flow) becomes large the resonances can impinge on the KH instability, pushing the critical wavenumber for direct instability to higher wavenumber $K$. This is of importance for understanding the growth of the KH instability at the surface of prominence threads or coronal loops because the oscillatory nature of the shear changes the onset of the KH instability. 

The general characteristics of the instabilities found in the model are:
\begin{enumerate}
\item {The frequency of a surface Alfv\'{e}n wave in the limit $\varepsilon \ll 1$ 
is modified by the oscillating flow,  
which is given in Eq.~(\ref{wave_KH}).
}

\item There exist surface Alfv\'{e}n waves that become resonant with the oscillatory driver as a result of the Doppler-shifting of their frequencies by the flow. {These waves undergo an exponential growth in amplitude. In the asymptotic limit of weak shear, the exponential growth associated with the strongest resonance can be calculated and is given in Eq.~\ref{max_parametric}.}:

\item Beyond this limit, the region in $K$-space where the parametric instability can grow approaches the KH unstable band, suppressing the KH instability in this region.

\item We expect that both these instabilities could exist in the solar atmosphere with characteristic time scales of $\sim 100$\,s, for wavelengths of $200$\,km around the flux tube.
\end{enumerate}

Returning to our initial motivation, though we use a highly simplified model when applied to the disruption of oscillating structures in the solar atmosphere, there are a number of conclusions we can draw.
{For an oscillating magnetic field in the solar atmosphere, the boundary conditions require modes to have wavenumbers along the magnetic field of $K_x>0$.
In fact, we calculate the minimum $K_x$ to be the point at which the surface Alfv\'{e}n wave, unmodified by the shear, would resonate with the driving frequency.
Due to the strengths of the magnetic field, most of the modes of interest would have to have the ratio of wavenumbers across and along the magnetic field of $K_y/K_x \ll 1$.}
One clear result of this study is that the instabilities that can develop from an oscillating flow are more complex than the KH instability of a non-oscillatory shear flow.
We can expect that resonances would play a role, and when the KH does grow, any vortices that are created will reverse the sign of their vorticity line with the change in sign of the vorticity of the forcing (in the solar case this is the large scale MHD wave).

The key point of this paper is that if we account for the oscillatory
nature of a wave in a flux tube, 
then we find that there are two types of instabilities, and that they can be excited for a wider range of wavenumbers (especially along the magnetic field) compared with the case of a constant shear. As such there are a richer array of disturbances on shorter scales along the magnetic field, and hence of ways to break down the original Alfv\'{e}nic wave, and possibly to disturb the underlying flux tube. An interesting extension to this work would be to perform a similar analysis on the surface of a flux tube similar to the model of \citet{SOLER2010} but considering an oscillating flow. Because the modes around the surface of a flux tube are quantised, as are the modes along a flux tube of finite length, this will make it harder to satisfy exact resonance, which could have an impact on the nature of the instability.

The existence of these two instabilities may have interesting consequences for the potential development of turbulence in the system under study. The non-linear development of the hydrodynamic KH instability can produce turbulence without a magnetic field.
In the MHD limit, there are extra complexities, but as shown in \citet{ANTOLIN2015} chaotic turbulent-like flows can develop.
The existence of parametric instabilities also has a connection to MHD turbulence, since the instability involves wave-interactions, which are also crucial for Alfv\'{e}nic turbulence \citep[e.g.][]{GOSR1995}.
It has been shown that the saturation of the parametric instability can have quite rich dynamics, giving rise to nonlinear oscillations, chaotic wave-wave interactions, and disordered wave turbulence \citep{WERS1980}. 
Therefore, the two MHD instabilities that can be expected to develop as a result of the oscillating shear flow we have studied potentially connect to the development of two different regimes of turbulence in an MHD system.


{The parametric instability can grow at larger scales across the magnetic field when the Alfv\'{e}nic Mach number $M_{\rm A}$ of the oscillating flow is increased or the density contrast between the two flow regions is decreased.
Dynamically this would be distinguished from a direct instability by the progressive increase in amplitude of a wave instead of the linear growth of a perturbation.
In the solar atmosphere, this density contrast is smaller in coronal loops than in prominence threads, making the parametric instability more likely to occur for coronal loop oscillations.
It is necessary to perform a range of MHD simulations to see if resonant enhancement of surface waves can happen at dynamically important scales under solar conditions.}
As the importance and the growth of the instability scales as $M_{\rm A}^2$, this could result in changes to the rate at which oscillations damp for increasing nonlinearity of the oscillation.
\citet{GODD2016} observed such a trend in coronal loops and it would be interesting to develop this connection.

\section*{Acknowledgements}

The authors would like to thank Prof. Andrew Gilbert for his comments which greatly improved the content of the manuscript.
AH is supported by his STFC Ernest Rutherford Fellowship grant number ST/L00397X/2 and by STFC grant ST/R000891/1.
AJB was supported by the Leverhulme Trust through the award of an Early Career Fellowship and by STFC Grant ST/R00059X/1.
IA acknowledges financial support from the Spanish Ministry of Economy and Competitiveness (MINECO) through projects AYA2014-55456-P (Bayesian Analysis of the Solar Corona), AYA2014-60476-P (Solar Magnetometry in the Era of Large Telescopes), from FEDER funds, and through a Ramon y Cajal fellowship.

\appendix
\section{Full Linear Stability Analysis}\label{lin_deriv}

Linearising about the basic state of an oscillatory shear flow, we expand each variable in the form $G=G_0+g$, where $G_0$ is a basic state variable and $g$ is its linear perturbation, we obtain the following set of equations:
\begin{align}
\frac{\partial\rho}{\partial t} +V_{y,0}\frac{\partial \rho}{\partial y}+v_z \frac{\partial \rho_0}{\partial z}=&0,\\
\rho_0 \frac{\partial \mathbf{v}}{\partial t}+\rho_0 V_{y,0}\frac{\partial \mathbf{v}}{\partial y}+\rho_0v_z\frac{\partial V_{y,0}}{\partial z}\mathbf{\hat{j}}=&-\nabla p +\mathbf{j}\times\mathbf{B}_0,\\
\frac{\partial \mathbf{b}}{\partial t}+V_{y,0}\frac{\partial \mathbf{b}}{\partial y}=&-b_z\frac{\partial V_{y,0}}{\partial z}\mathbf{\hat{j}}+B\frac{\partial \mathbf{v}}{\partial x},\\
\nabla \cdot \mathbf{v}=&0,\\
\nabla \cdot \mathbf{b}=&0,
\end{align}
where $\mathbf{\hat{j}}$ is the unit vector in the y direction.
We assume normal modes of the form $f(x,y,z,t)=\tilde{f}(z,t)exp(ik_xx+ik_yy)$, and obtain the following system of linear equations:
\begin{align}
D\rho +\tilde{v}_z \frac{\partial \rho_0}{\partial z}=&0,\\
\rho_0 D\tilde{v}_x=&-ik_x \tilde{p}, \label{x_vel}\\
\rho_0 D\tilde{v}_y+\rho_0\tilde{v}_z\frac{\partial V_{y,0}}{\partial z}=&-ik_y \tilde{p} +\frac{B}{4\pi}(ik_x\tilde{b}_y-ik_y\tilde{b}_x),\label{y_vel}\\
\rho_0 D\tilde{v}_z=&-\frac{\partial \tilde{p}}{\partial z} -\frac{B}{4\pi}\left(\frac{\partial\tilde{b}_x}{\partial z}-ik_x\tilde{b}_z\right),\label{z_vel}\\
D \tilde{b}_x=&ik_xB \tilde{v}_x,\label{induction1}\\
D \tilde{b}_y+b_z\frac{\partial V_{y,0}}{\partial z}=&ik_xB \tilde{v}_y,\label{induction2}\\
D \tilde{b}_z=&ik_xB \tilde{v}_z,\label{induction3}\\
ik_x\tilde{v}_x+ik_y\tilde{v}_y=&-\frac{\partial \tilde{v}_z}{\partial z},\label{incomp}\\
ik_x\tilde{b}_x+ik_y\tilde{b}_y=&-\frac{\partial \tilde{b}_z}{\partial z},\label{div_b},
\end{align}
where $D=\partial/\partial t + ik_yV_{y,0}$, i.e. the advective derivative.

The first step is to take the $x$ derivative of Eq.~(\ref{x_vel}) and add this to the $y$-derivative of Eq.~(\ref{y_vel}) and in conjunction with Eqs.~(\ref{incomp}) and (\ref{div_b}) gives:
\begin{equation}
-\rho_0 D\frac{\partial \tilde{v}_z}{\partial z}+\rho_0\tilde{v}_z\frac{\partial V_{y,0}}{\partial z}=k^2 \tilde{p} -\frac{ik_xB}{4\pi}\frac{\partial \tilde{b}_z}{\partial z}+\frac{k^2B}{4\pi}\tilde{b}_x\label{composite}.
\end{equation}
From here we can introduce the Lagrangian displacement $\tilde{\boldsymbol{\eta}}$ defined by
\begin{align}
\tilde{v}_x=&D\tilde{\eta}_x,\\
\tilde{v}_z=&D\tilde{\eta}_z,
\end{align}
which implies that 
\begin{align}
\tilde{b}_x=&ik_xB\tilde{\eta}_x,\\
\tilde{b}_z=&ikxB\tilde{\eta}_z.
\end{align}
Substituting these into Eqs.~(\ref{z_vel}) and (\ref{composite}) leads to:
\begin{align}
-\rho_0 D^2\frac{\partial \tilde{\eta}_z}{\partial z}=&k^2 \tilde{p} +\frac{k_x^2B}{4\pi}\frac{\partial \tilde{\eta}_z}{\partial z}+ik_x\frac{k^2B^2}{4\pi}\tilde{\eta}_x,\\
\rho_0 D^2\tilde{\eta}_z=&-\frac{\partial \tilde{p}}{\partial z} -ik_x\frac{B^2}{4\pi}\left(\frac{\partial\tilde{\eta}_x}{\partial z}-ik_x\tilde{\eta}_z\right).\label{new_z_comp}
\end{align}
It is simple to show that:
\begin{equation}
-\frac{\partial\tilde{p}}{\partial z}=\frac{1}{k^2} \frac{\partial}{\partial z}\left(\frac{k_x^2B}{4\pi}\frac{\partial \tilde{\eta}_z}{\partial z}+ik_x\frac{k^2B^2}{4\pi}\tilde{\eta}_x+\rho_0 D^2\frac{\partial \tilde{\eta}_z}{\partial z}\right),
\end{equation}
which substituted into Eq.~(\ref{new_z_comp}) gives:
\begin{equation}
\frac{\partial}{\partial z}\left(\left(\rho_0 D^2+\frac{k_x^2B}{4\pi}\right)\frac{\partial \tilde{\eta}_z}{\partial z}\right) -k^2\left(\rho_0 D^2+\frac{k_x^2B^2}{4\pi}\right)\tilde{\eta}_z=0.\label{eta_z_equation}
\end{equation}

Looking at Eq.~(\ref{eta_z_equation}) at $z\ne0$ gives:
\begin{equation}
\left(\rho_0 D^2+\frac{k_x^2B}{4\pi}\right)\left(\frac{\partial^2 }{\partial z^2}-k^2\right)\tilde{\eta}_z=0.
\end{equation}
We know that at the $z$-dependence of the solution is of the form $\exp(-k|z|)$, which is the same as in the MHD KH instability with constant shear \citep{CHAN1961}.
Using the continuity of $\tilde{\eta}_z$ at the boundary, we have that:
\begin{equation} 
\tilde{\eta}_z(z,t)=\eta(t)\exp(-k|z|).
\end{equation}

Equation~(\ref{eta_z_equation}) can be integrated over the discontinuity to give:
\begin{equation}
\left(\rho_0 D_+^2+\frac{k_x^2B}{4\pi}\right)\eta+\left(\rho_0 D_-^2+\frac{k_x^2B}{4\pi}\right)\eta=0,
\end{equation}
or, in full, 
\begin{align}
\frac{d^2 \eta}{d t^2}&+2ik_y(\alpha_+V_++\alpha_-V_-)\frac{d \eta}{d t}+\left[ik_y\left(\alpha_+\frac{d V_+}{d t}+\alpha_-\frac{d V_-}{d t} \right)\right. \nonumber\\& \left.-k_y^2(\alpha_+V_+^2+\alpha_-V_-^2) +\frac{k_x^2B^2}{2\pi (\rho_++\rho_-)}\right]\eta=0,
\end{align}
where $\alpha_{\pm}=\rho_{\pm}/(\rho_++\rho_-)$.

If we define $\eta=\hat{\eta}\exp(-ik_y\int\alpha_+V_++\alpha_-V_-dt)$, we can remove the term with the first time derivative to obtain:
\begin{equation}
\frac{d^2 \hat{\eta}}{d t^2} +\left[\frac{k_x^2B^2}{2\pi (\rho_++\rho_-)} -k_y^2\alpha_+\alpha_-(V_+-V_-)^2 \right]\hat{\eta}=0.
\end{equation}
By definition, $V_+-V_-=\Delta V_0 \cos \omega_0 t$, therefore $(V_+-V_-)^2=\Delta V_0^2( \cos (2\omega_0 t) +1)/2$, so we have
\begin{equation}
\frac{d^2 \hat{\eta}}{d t^2} +\left[\frac{k_x^2B^2}{2\pi (\rho_++\rho_-)} -\frac{1}{2}k_y^2\alpha_+\alpha_-\Delta V_0^2\left(1+\cos (2 \omega_0 t) \right)\right]\hat{\eta}=0.
\end{equation}
This is in the form of a Mathieu equation, and as such its properties (both the instability and wave solutions) can be easily understood.

\bsp	
\label{lastpage}
\end{document}